\documentclass[conference]{IEEEtran}
\IEEEoverridecommandlockouts

\usepackage{amsmath,amssymb,amsfonts}
\usepackage{comment}
\usepackage{ulem}
\usepackage{cite}
\usepackage{framed}
\usepackage{color}
\usepackage{slashbox}
\usepackage{listings}
\usepackage{latexsym}
\usepackage{graphicx}
\usepackage{url}

\usepackage{algorithm}
\usepackage{algpseudocode}

\newtheorem{defdef} {Definition}
\begin{document}

\title{Unsupposable Test-data Generation for Machine-learned Software
}

\author{\IEEEauthorblockN{Naoto Sato, Hironobu Kuruma, and Hideto Ogawa}
\IEEEauthorblockA{\textit{Research \& Development Group, } 
\textit{Hitachi, Ltd.}\\
naoto.sato.je@hitachi.com}

}

\maketitle

\begin{abstract}
As for software development by machine learning, a trained model is evaluated by using part of an existing dataset as test data. 
However, if data with characteristics that differ from the existing data is input, the model does not always behave as expected. Accordingly, to confirm the behavior of the model more strictly, it is necessary to create data that differs from the existing data and test the model with that different data. 
The data to be tested includes not only data that developers can suppose ({\it supposable} data) but also data they cannot suppose ({\it unsupposable} data). To confirm the behavior of the model strictly, it is important to create as much unsupposable data as possible. In this study, therefore, a method called ``unsupposable test-data generation" (UTG)---for giving suggestions for unsupposable data to model developers and testers---is proposed. UTG uses a variational autoencoder (VAE) to generate unsupposable data. The unsupposable data is generated by acquiring latent values with low occurrence probability in the prior distribution of the VAE and inputting the acquired latent values into the decoder. 
If unsupposable data is included in the data generated by the decoder, the developer can recognize new unsupposable features by referring to the data. On the basis of those unsupposable features, the developer will be able to create other unsupposable data with the same features. The proposed UTG was applied to the MNIST dataset and the House Sales Price dataset. The results demonstrate the feasibility of UTG. 
\end{abstract}


\section{Introduction}\label{intro}
In recent years, software developed by machine learning has been introduced in various fields of industry. In conjunction with this trend, techniques for checking the behavior of such software developed by machine learning have been developed. Some examples of proposed techniques 
verify deep neural networks (DNNs) and ensemble models using decision trees to determine whether their input data and output data satisfy certain properties \cite{reluplex}\cite{dlv}\cite{starbased}\cite{xgboostveri}. Hereafter, software developed by machine learning (that is, trained software) is referred to as a {\it model} in this paper. By using those verification techniques, it is possible to verify  exhaustively whether a model satisfies requirements regarding safety, for example. However, it is known that exhaustive verification cannot be completed within a practical time when the scale of the model is large
Moreover, even if the correctness of the model cannot be defined as a property, these exhaustive verification techniques cannot be used. For example, in the case of an image-recognition problem, the requirement that ``A given image should be correctly identified as a person." cannot be defined as a property. In such cases, verifying the model by testing is effective. 

As for development of a model, a certain part of an existing dataset is used as training data, and the rest is used as test data for evaluating the trained model. The generalizability of the model can be confirmed by using different data from the training data as the test data. With this method, it is possible to check that the developed model behaves as expected with existing data; however, if data with different characteristics from the existing data is input, the behavior does not always turn out as expected. Therefore, to confirm the behavior of the model more strictly, it is necessary to create input data differing from the existing data and test the model with that different data. However, even in the case of data differing from existing data, data that does not need to be supposed is excluded from the test data. This supposition-unrequired data is not always outside the domain of the input data. For example, in the case of a model that performs image recognition, 
the model accepts all images of a certain size as input data; however, the data that actually needs to be supposed is only a part of those images, that is, 
images of things that exist in the real world. Hereafter, the data that does not need to be supposed is called {\it unreal} data. The boundary between real data (which needs to be supposed) and unreal data is not always clear. For example, in the case of an image-recognition model, the data that needs to be supposed comprises images of things that exist in the real world; however, it is difficult to define exactly what kind of images they are. 

Furthermore, data that must be supposed can be classified as data that model developers and testers can easily suppose and data that is difficult to suppose. Hereafter, the former is called {\it supposable} data and the latter is called {\it unsupposable} data. Existing data is one part of supposable data. Whether certain data is supposable or unsupposable depends on the person doing the supposing, so the boundaries between them are ambiguous. In the following, the person making suppositions about input data is referred to as the {\it developer}. 

\begin{figure}[htbp]
\begin{center}
\scalebox{0.40}{\includegraphics[bb=0 0 585 300]{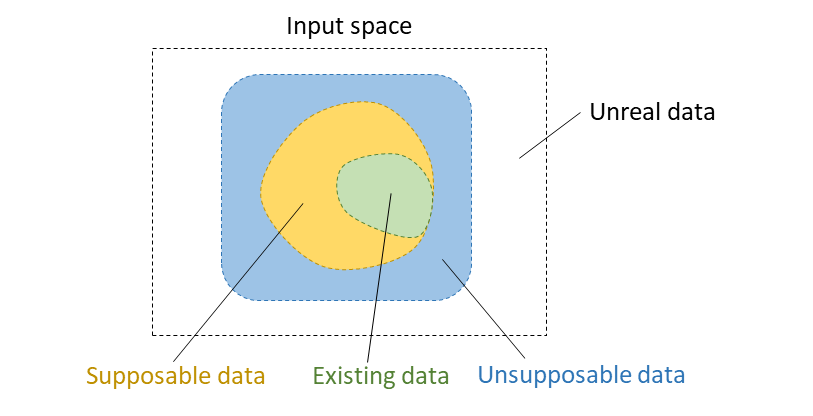}}
\end{center}
\caption{Relationship between existing data, unreal data, supposable data, and unsupposable data}
\label{fig01}
\end{figure}

From the above discussion, it can be said that although the boundaries between them are ambiguous, data can be classified as three types: supposable data, unsupposable data, and unreal data (Fig. \ref{fig01}). The data to be tested are supposable data and unsupposable data, of which the supposable data can be easily created. Therefore, to confirm the behavior of a model strictly, it is preferable to create as much unsupposable data as possible. Therefore, we propose ``unsupposable test-data generation" (UTG) as a technique aiming to give suggestions for unsupposable data to developers.
UTG uses a variational autoencoder (VAE) to generate data. As for a normal VAE, a latent value is sampled according to the prior distribution $P$, and data is generated by decoding the sampled latent value. Since the data generated in this way has similar characteristics to the existing data used for training the VAE, it is likely to be supposable data. On the other hand, with UTG, a latent value with a low probability of occurrence in $P$ is acquired and decoded. Consequently, data with different characteristics from the existing data is generated. As described above, however, the boundaries between unreal data, supposable data, and unsupposable data are ambiguous, so the data generated by UTG is not always unsupposable data. Accordingly, UTG has parameters which are used to change the rarity of the acquired latent value.
The developer changes the values of the parameters while referring to the generated data and exploratively determines the values of the parameters 
so that as much unsupposable data as possible is included in the data generated by the decoder.
If unsupposable data is included in the data generated by the decoder, the developer can recognize unsupposable features by referring to that data. On the basis of those features, the developer can create other unsupposable data with those unsupposable features. In this study, by applying UTG to the MNIST dataset \cite{mnist} and the House Sales Price dataset, it was confirmed that it is possible to generate unsupposable data from data generated by UTG. The first contribution of this paper is to propose UTG and show its implementation.
The second contribution is to demonstrate the feasibility of UTG through case studies.

The rest of this paper is organized as follows. In Section \ref{relwork}, related research is described, and the position and aim of UTG are clarified. In Section \ref{background}, the variational auto-encoder (VAE) and the VQ (vector quantized)-VAE with PixelCNN, which are the basis for implementing UTG, are described. In Section \ref{utg}, the UTG concept is explained, and the means of implementing UTG using VAE and VQ-VAE with PixelCNN is described. In Section \ref{experiment}, the results of the case study are presented, and in Section \ref{discuss}, those results are evaluated and discussed. In Section \ref{conclusion}, the conclusions of this study are given, and future work is discussed.

\section{Related Work}\label{relwork}
As approaches to create new data for testing a model, several methods that focus on the activation status of the neurons that make up a DNN are known \cite{deepxplore} \cite{deeptest}\cite{tensorfuzz}. As for these methods, new data is created by processing existing data so as to activate neurons that were not activated when the existing data was input. For example, in the case of image data, processing such as increasing the brightness 
 or adding white noise is performed. Then, both the processed data and the unprocessed data are input to the model, and matching of the output values is confirmed. It has been shown experimentally that this method can demonstrate incorrect behavior of the model. A test method that inputs the data before and after processing into the model and compares and evaluates the output values in this way is called metamorphic testing \cite{metamor1}\cite{metamor2}. In an example in which metamorphic testing is applied to an objective-perception model of self-driving cars, small particles in the air and noise from sensors are given as effects \cite{driverless}. Moreover, adding perturbation---instead of the semantic effects described above---to the extent that the original image data does not change significantly is also a way of processing \cite{adversarial}.


The means of processing of the input data, such as increasing brightness, adding small particles, or adding perturbation, is defined according to the problem that the model solves. 
It is necessary to select how to processing existing data so that the processed data becomes the data that should be supposed as the input data, that is, real data. For example, in the case of a model used in an environment in which brightness is strictly controlled, it is not necessary to suppose a change in brightness. On the other hand, in the case of a model that is used outdoors day and night, it is necessary to assume a change in brightness. The data that can be created by these methods is supposable data because it is created by processing existing data in a way that can be supposed by the developer


Moreover, using GAN \cite{gan1}\cite{gan2} or VAE \cite{vae1}\cite{vae2}\cite{vae3}, a natural image can be generated by changing some of the attributes from an existing image \cite{unit}\cite{cyclegan} \cite{deeproad}. As for these methods, a neural network is trained by using a set of images having one attribute (A) and another set of images having another attribute (B). Then, when an arbitrary image with attribute A is input to the trained neural network, attribute A is deleted from that image, and an image with added attribute B is output instead. To generate data using these methods, the developer needs to define the attributes of the data to be generated. That is, the data that can be generated by these methods is taken as supposable data because it has the attributes that are expected by the developer. On the other hand, as for UTG, unsupposable data (which is difficult for the developer to expect) is generated without inputting a supposition by the developer such as the data-processing way, attributes, and so on.


\section{Background}\label{background}

\subsection{Variational Auto-encoder}\label{sec_vae}
\begin{figure}[htbp]
\begin{center}
\scalebox{0.38}{\includegraphics[bb=0 0 650 95]{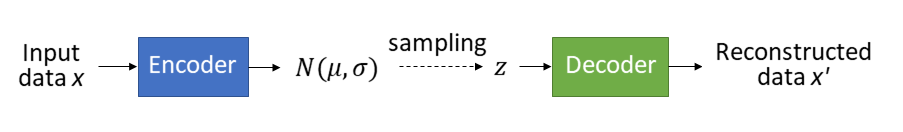}}
\end{center}
\caption{Structure of VAE}
\label{fig02}
\end{figure}

\begin{figure}[htbp]
\begin{center}
\scalebox{0.38}{\includegraphics[bb=0 0 590 125]{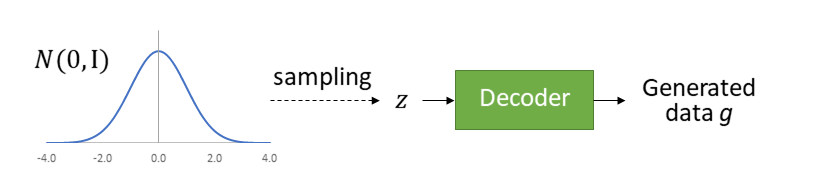}}
\end{center}
\caption{Generating data by VAE}
\label{fig03}
\end{figure}

A variational autoencoder (VAE) is composed of an encoder and a decoder. From $z$, which represents an unobservable feature of input data $x$, the decoder creates input data $x'$ that may correspond to $z$. Here, $z$ is called a latent variable and holds a $K$-dimensional vector value. Hereafter, the prior distribution of $z$ is represented by $P$, which has a mean of zero, and the variance-covariance matrix is assumed to be a Gaussian distribution with identity matrix $I$. From input data $x$, the encoder creates a distribution of values of $z$ from which $x$ could have been generated.  The distribution of $z$ created by the encoder is assumed to be Gaussian. Here, the case in which $x0$ (corresponding to a certain $z0$) is given to the encoder is considered. $z1$ obtained by sampling the Gaussian distribution output by the encoder is expected to be a value close to $z0$. The value obtained by giving $z1$ to the decoder is taken as $x1$, which is expected to be similar to $x0$. To meet these expectations, the encoder and decoder are trained with the structure shown in Fig. \ref{fig02}. Hereafter, $T$ represents the existing dataset that is used for training.
The encoder accepts input data $x$ and outputs mean $ \mu $ and variance-covariance matrix $ \sigma $, which are parameters of the Gaussian distribution. And $z$ is obtained by sampling from the distribution determined by $ \mu $ and $ \sigma $. (Actually, instead of sampling, a method called reparameterization trick is used to calculate $z$ from $ \mu $ and $ \sigma $ by using random noise \cite{vae1}.) The decoder accepts $z$ and outputs reconstructed data $x'$. The trained decoder is used for generating data (Fig. \ref{fig03}).
The value of $z$ is obtained by sampling the prior distribution $P$ of $z$, and data $g$ is generated by giving that value to the trained decoder. The distribution of $g$ is similar to the distribution of $T$, so the decoder is used to generate data similar to $T$.

\subsection{VQ-VAE with PixelCNN}\label{sec_vqvae}
\begin{figure}[htbp]
\begin{center}
\scalebox{0.38}{\includegraphics[bb=0 0 662 136]{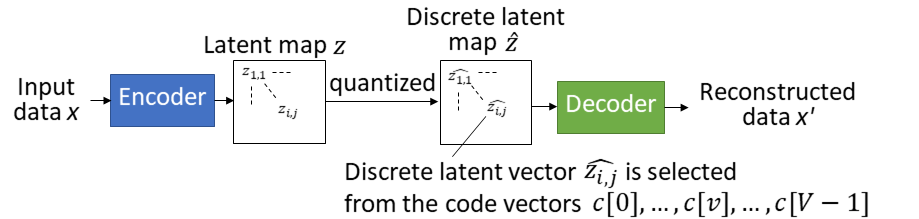}}
\end{center}
\caption{Structure of VQ-VAE}
\label{fig04}
\end{figure}

When the target is image data, a VQ (vector quantized)-VAE \cite{vqvae} can be used to generate data with better image quality than that possible with a VAE. The structure of VQ-VAE is shown in Fig. \ref{fig04}.
The VQ-VAE encoder accepts input data $x$ and outputs $z$, which represents a $ i \times j $ matrix whose elements are latent vector $z_{i, j}(1 \leq i \leq I, 1 \leq j \leq J)$. Latent vector $z_{i, j}$ forming $z$ is a $K$-dimensional vector. 
$z_{i, j}$ that composes $z$ is replaced with a vector of fixed values included in a list called a codebook. The latent vector obtained after replacement is called a discrete latent vector, which is represented as $ \hat{z_{i, j}} $. The number of fixed value vectors included in the codebook---denoted by $V$---is finite. In this paper, the fixed-value vectors in the codebook are called {\it code vector} which are represented as $c[0], ..., c[v],..., c[V-1]$. Among the code vectors, the one having the closest Euclidean distance to $z_{i,j}$ is selected as $ \hat{z_{i,j}} $. As a result, latent vector $z_{i,j}$ is ``quantized" to a discrete latent vector given as $ \hat{z_{i,j}} $.

\begin{figure}[htbp]
\begin{center}
\scalebox{0.38}{\includegraphics[bb=0 0 719 163]{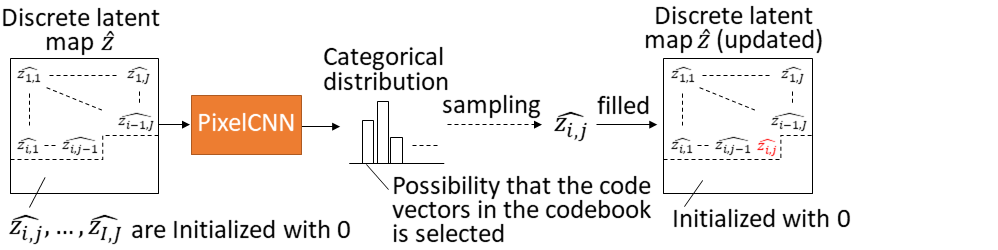}}
\end{center}
\caption{PixelCNN over latents}
\label{fig05}
\end{figure}

As for VQ-VAE, an autoregressive model is used to estimate the prior distribution $P$ and generate data. PixcelCNN is cited by Oord, et al. \cite{vqvae} as an example of an autoregressive model for images, and they cite WaveNet as an example of an autoregressive model for audio. In this paper, a method for generating image data using PixelCNN is shown in Fig. \ref {fig05}.

PixelCNN accepts discrete latent map $ \hat{z} $ in which values from $ \hat{z_{1,1}} $ to $ \hat{z_{i, j-1}} $ were filled as input. The categorical distribution for determining the value of $ \hat{z_{i,j}} $ is then output. As described above, $ \hat{z_{i,j}} $ is one of the code vectors included in the codebook. The categorical distribution output by PixelCNN represents the probability that each code vector included in the codebook will be selected as $ \hat{z_{i,j}} $, the value of which is determined by sampling from the categorical distribution. When existing dataset $T$ is input into the trained VQ-VAE encoder, discrete latent maps corresponding to $T$ are obtained. These discrete latent maps ate then used to train PixelCNN. 

\begin{figure}[htbp]
\begin{center}
\scalebox{0.38}{\includegraphics[bb=0 0 816 346]{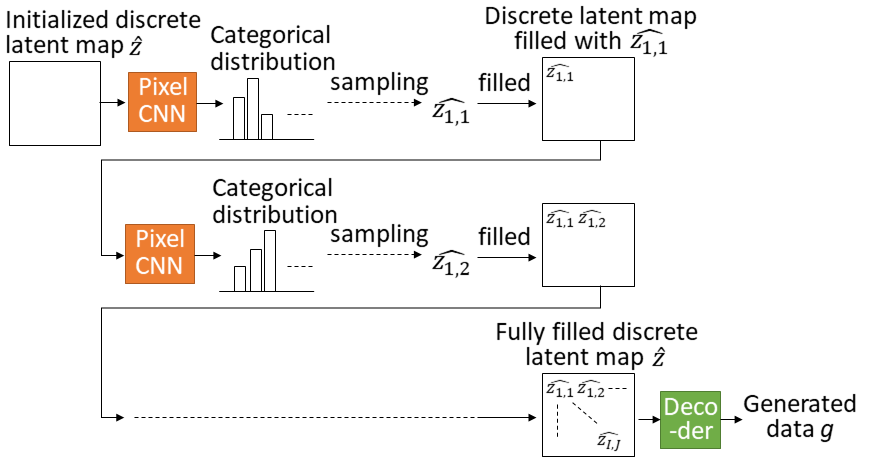}}
\end{center}
\caption{Generating data by VQ-VAE with PixelCNN}
\label{fig06}
\end{figure}

As shown in Fig. \ref{fig06}, a discrete latent map can be generated by using the trained PixelCNN recursively. Then, the generated discrete latent map is input into the VQ-VAE decoder to generate data $g$. It is known that using VQ-VAE and PixelCNN makes it possible to generate image data with better image quality than that possible with VAE \cite{vqvae}. In particular, it is expected to prevent blurring of contours.

\section{Unsupposable Test-data Generation(UTG)}\label{utg}

\subsection{Concept}\label{concept}
As for UTG, unsupposable data is generated by acquiring a value with a low occurrence probability in prior distribution $P$ of latent variable $z$. As described in Section \ref{sec_vae}, $P$ 
 is given as $ \mathcal{N}(0, I)$
As for usual data generation using a decoder, generated data $g$ is obtained by sampling the value of $z$ according to distribution $P$ and giving it to the trained decoder. The dataset generated by the decoder is denoted by $G$ ($g \in G$). The distribution of $G$ is similar to that of existing dataset $T$ used for training because the encoder and decoder are optimized for $T$. Here, it is assumed that the value of $z$ obtained from $P$ has a high occurrence probability, that is, 
its absolute deviation in $P$ is small.
Since $z$ represents an unobservable feature of $x$, it can be said that $g$ generated from $z$ is likely to have features with high occurrence probability in $T$. On the contrary, when $z$ obtained from $P$ has a low occurrence probability, it can be said that the generated $g$ is likely to have features which are not commonly observed in $T$.
In other words, $g$ might contain unsupposable features.

Utilizing this fact, UTG intentionally acquires a latent value with a low occurrence probability in $P$, which is denoted $z^{u}$. The acquired latent value $z^{u}$ is input into the decoder to generate data $g^{u}$. It is considered that the dataset generated by this method is more likely to contain unsupposable data than one generated from $z$ obtained according to $P$. The dataset generated by this method is hereafter called a {\it likely-unsupposable} dataset $LU$. By referring to $LU$, a developer can get suggestions for unsupposable data. However, as described in Section \ref{intro}, the boundaries between supposable data, unsupposable data, and unreal data are ambiguous and depend on 
the developer, so this method cannot always generate unsupposable data. Even if the boundaries between supposable data, unsupposable data, and unreal data are known, the lowest probability of occurrence in $P$ that can be adopted as $z^{u}$ to generate unsupposable data also depends on the results of training the VAE. Therefore, as for UTG, the method of obtaining $z^{u}$ from $P$ is parameterized, and the rarity of $z^{u}$ to be obtained can be changed by changing the values of the parameters. The developer changes the values of the parameters while referring to the generated dataset $LU$ and adjusts them so that $LU$ contains a lot more unsupposable data. By adjusting the occurrence probability of the acquired $z^{u}$ in $P$ in this manner, the possibility of generating unsupposable data is improved exploratively.

\subsection{Implementation on VAE}\label{impvae}
\begin{figure}[htbp]
\begin{center}
\scalebox{0.7}{\includegraphics[bb=0 0 329 234]{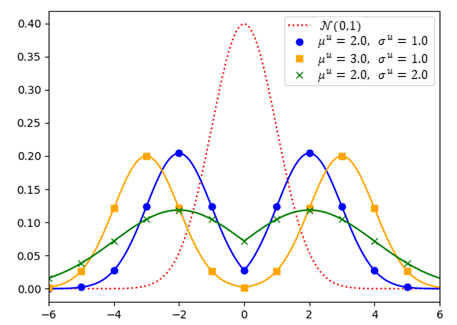}}
\end{center}
\caption{Probability density functions $f$ of $z^{u}[k]$}
\label{fig07}
\end{figure}

The concept of UTG described in Section \ref{concept} is to (i) ``acquire $z^{u}$ with low occurrence probability in priority distribution $P$ and generate $g^{u}$" and (ii) ``to exploratively adjust the occurrence probability in $P$ of $z^{u}$ 
by changing the values of paremeters." Based on these two concepts, the probability density function $f$ that gives the probability distribution of $z^{u}[k] (0 \leq k \leq K-1)$, which is an element of $z^{u}$, is defined as follows:
\begin{defdef}
\label{def01}
\begin{eqnarray*}
f(z^{u}[k])_{\mu ^{u}, \sigma ^{u}} = \begin{cases}
 \frac{1}{A} Normal(z^{u}[k], | \mu^{u} |, \sigma^{u}) & (z^{u}[k] \geq 0) \\
 \frac{1}{A} Normal(z^{u}[k], - | \mu^{u} |, \sigma^{u}) & (z^{u}[k] < 0) \\
\end{cases}
\end{eqnarray*}
\end{defdef}
where $ \mu^{u} $ and $ \sigma^{u} $ are parameters for adjusting the occurrence probability of $z^{u}$ in $P$. $Normal$ is the probability density function of a normal distribution, that is, $Normal(x, \mu, \sigma) = \frac{1}{\sqrt{2 \pi \sigma^2}} exp \left( - \frac{{(x - \mu)}^2}{2 {\sigma}^2}  \right)$. And $A$ is a normalizing constant defined as follows: 
\begin{defdef}
\label{def02}
\begin{eqnarray*}
A =2 \left( (1 / 2) + \int_{0}^{ | \mu^{u} | } Normal(x, | \mu^{u} |, {\sigma^{u}}) dx \right)
\end{eqnarray*}
\end{defdef}
In Fig. \ref {fig07}, functions $f$ when $ \mu^{u} $ and $ \sigma^{u}$ are changed are shown as solid lines. For comparison, $P = \mathcal{N}(0, 1)$ is shown as a dotted line.
As shown in Fig. \ref{fig07}, it is highly likely that the occurrence probability of $z^{u}[k]$ obtained from the distribution given by $f$ is low (but not too low) at $P = \mathcal{N}(0, 1)$. Moreover, by changing the values of $ \mu^{u} $ and $ \sigma^{u} $, it is possible to adjust the occurrence probability of $z^{u}[k]$ in $P$. To sample $z^{u}[k]$ from $f$, Markov Chain Monte Carlo methods can be used. UTG adopts the Metropolis algorithm \cite{metropolis}. Likely-unsupposable dataset $LU$ is generated by inputting $z^{u} = \{ z^{u}[0], ..., z^{u}[K-1] \}$ to the decoder of the trained VAE.

\subsection{Implementation on VQ-VAE with PixelCNN}\label{impvq}
\begin{figure*}[htbp]
\begin{center}
\scalebox{0.4}{\includegraphics[bb=0 0 955 335]{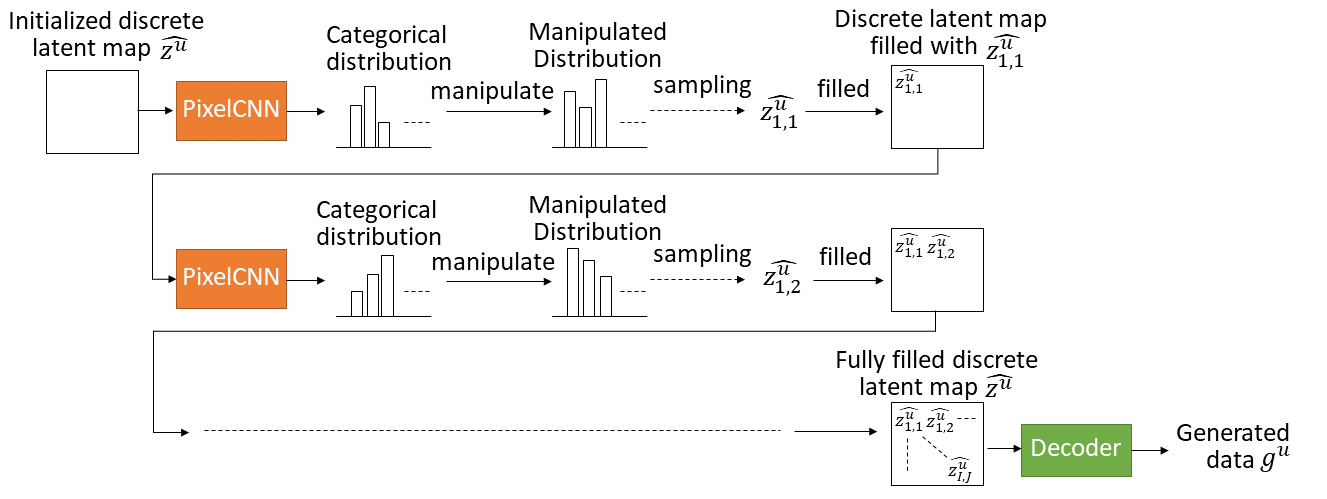}}
\end{center}
\caption{Implementation of UTG on VQ-VAE with PixelCNN}
\label{fig08}
\end{figure*}

As UTG for image data, implementing UTG by using VQ-VAE with PixelCNN is proposed. By implementing UTG on VQ-VAE with PixelCNN, it is possible to generate unsupposable data with better image quality than that possible with VAE. On the contrary, VQ-VAE with PixelCNN cannot be applied to structured data, so when UTG is applied to structured data, the implementation of UTG with VAE is used. As shown in Section \ref{sec_vqvae}, in the case of VQ-VAE with PixelCNN, PixelCNN is used to estimate prior distribution $P$. Specifically, a discrete latent map $ \hat{z} $ is acquired on the basis of the categorical distribution output by PixelCNN. Therefore, by manipulating this categorical distribution
, $ \hat{z^{u}} $ with a low occurrence probability in $P$ is obtained (Fig. \ref{fig08}).
The categorical distribution output by PixelCNN is given as $d$, which represents the probability that code vector $c[0],..., c[v], ..., c[V-1]$ contained in the codebook will be selected as $ \hat {z_{i,j}} $. Thus, $d$ is expressed as $d= \{ d[0], ..., d[v], ..., d[V-1] \} ( \sum_{0 \leq v \leq V-1}d[v] = 1 )$. Various possible methods for manipulating $d$ so as to acquire $z^{u}$ with low occurrence probability in $P$ are available; however, hereafter, the following algorithm \ref{al01} is adopted:
\begin{algorithm}[htb]
\caption{Algorithm for manipulating categorical distribution $d$}
\label{al01}
\begin{algorithmic}[1]
\Require $d$, $t$
\Ensure $d^{u}$
\newline

\State $d^{u}$ $\leftarrow$ $d$
\State $sum$ $\leftarrow$ $0$
\For{each $v$ in $ \{0, ..., V-1 \} $ }
    \If{ $d[v] > t $ }
        \State $diff$ $\leftarrow$ $d[v] - t$
        \State $sum$ $\leftarrow$ $sum + diff$
        \State $d^{u}[v]$ $\leftarrow$ $t$
    \EndIf
\EndFor
\For{each $v$ in $ \{0, ..., V-1 \} $ }
    \State $d^{u}[v]$ $\leftarrow$ $d^{u}[v] + ( sum / V) $
\EndFor
\State \Return $d^{u}$
\end{algorithmic}
\end{algorithm}

As for this algorithm, when $d[v] (0 \leq v \leq V-1)$ exceeds threshold $t$, the value of $d[v]$ is reduced to $t$. Then, the values of $diff$ for which $d[v]$ exceeds $t$ are totaled and evenly distributed to each $d[v]$. With this algorithm, the code vector that has low probability of being selected from the original $d$ is selected with higher probability. $ \hat{z^{u}} $ is created by using categorical distribution $d^{u}$ provided in this algorithm, and it is input into the trained decoder to generate likely-unsupposable dataset $LU$.

\section{Experiment}\label{experiment}

\subsection{House Sales Price Dataset}\label{house}
UTG was applied to the dataset \footnote {https://www.kaggle.com/harlfoxem/housesalesprediction} used for the developing a housing-price-forecast model. This dataset consists of 18 attributes (excluding price), and of those attributes, 14 attributes (such as number of bedrooms, number of bathrooms, and size of living room) were selected and taken as existing dataset $T$. Since this data is not image data but structured data, the VAE implementation described in Section \ref{impvae} was applied. The value of parameter $(\mu^{u}, \sigma^{u})$ was determined by trial and error to be $(5, 5)$. Here, trial and error means the following: the value of $(\mu^{u}, \sigma^{u})$ is tentatively set, and $LU$ generated as a result is referred to. The developer changes the value of $(\mu^{u}, \sigma^{u})$ several times so as to maximizes the number of unsupposable data in generated $LU$. 
In that referential way, the value of $(\mu^{u}, \sigma^{u}$ is determined. By acquiring $100$ values of $z^{u}$ and inputting them into the decoder, a likely-unsupposable dataset $LU$ consisting of $100$ items of $g^{u}$ was generated. Of the element values of generated data $g^{u}$, the values given as integers or categorical values are rounded. Then, for all items of data $g^{u}$, whether data with similar characteristics to $g^{u}$ is included in the existing dataset $T$ and whether $g^{u}$ is real data were judged manually. If $T$ does not contain data with similar characteristics to $g^{u}$ and if $g^{u}$ is real data, $g^{u}$ is possibly unsupposable data. However, whether the data is actually unsupposable depends on the developer. In this way, data that may be taken as unsupposable data depending on the developer is referred to as {\it unsupposable-data candidate} hereafter. It was confirmed that at least $32$ unsupposable-data candidates (shown in Table \ref{table01}) were included in the generated $LU$.
\begin{table*}[htb]
\begin{center}
\scalebox{1.0}{\includegraphics[bb=0 0 722 255]{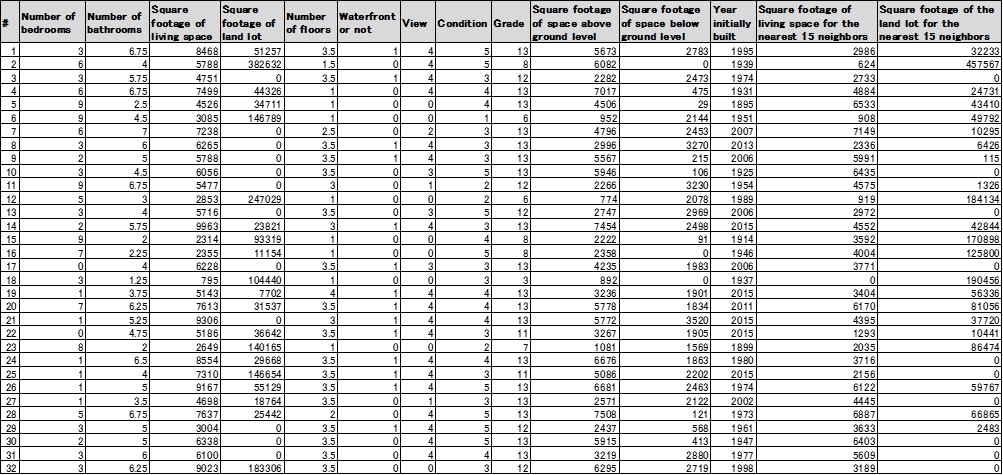}}
\end{center}
\caption{Unsupposable data in House Sales Price dataset \label{table01}}
\end{table*}

For example, data \#1 has 6.75 bathrooms and 3.5 floors, and data which has similar values is not included in existing dataset $T$. Moreover, other data with more floors (e.g., 3.5 floors) than bathrooms (e.g., 2.5 bathrooms) and data with more bathrooms than floors (e.g., 8 bathrooms and 2.5 floors) is included in $T$. Data \#1 is considered to be real because it seems that the number of bathrooms and floors is better balanced than the other data. Similarly, as for data \#2, the living area is about 6,000 square feet, and the average square footage of land lot of 15 neighbors is about 460,000 square feet; however, similar data is not included in $T$. Data for the square footage of 15 lots of the nearest neighbors is about 9000 square feet with the size of the living area of about 12,000 square feet and data for the average square footage of 15 lots of the nearest neighbors is 870,000 square feet with the size of the living area of about 6,000 square feet, are included in $T$; therefore, data \#2, which has better balance than these existing data, is considered to be real data. From the above results, it can be said data \#1 and \#2 are unsupposable-data candidates. Similarly, it is affirmed that the other data in Table \ref{table01} are real, but they have characteristics that are not included in $T$.

\subsection{MNIST Dataset}
\begin{figure}[htbp]
\begin{center}
\scalebox{0.7}{\includegraphics[bb=0 0 329 251]{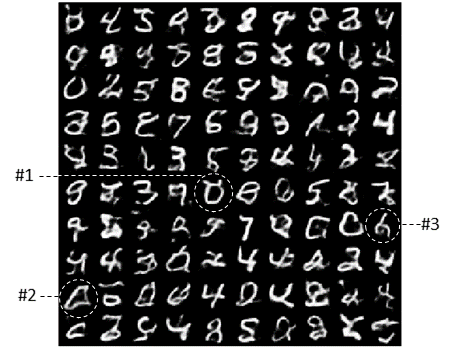}}
\end{center}
\caption{Example of data generated with MNIST dataset}
\label{fig09}
\end{figure}

UTG was applied with the MNIST dataset (used for developing digit-recognition models) as existing dataset $T$. Since this dataset contains image data, the implementation of UTG using VQ-VAE with PixelCNN (described in Section \ref{impvq}) was applied. Similar to Section \ref{house}, the value of parameter $t$ was changed by trial and error and finally set to $t=0.6$. Generated $LU$ (consisting of $100$ images) is shown in Fig. \ref{fig09}. 

As for image \#1 in Fig. \ref{fig09}, it looks like a zero with short lines like rabbit's ears, and it is affirmed that images with similar characteristics are not included in $T$. Similarly, it is affirmed that a zero image \#2---with a sharp point on its right side---is not included in $T$. Moreover, image \#3 has a point at its center of the circle of the six and the bottom line is faint and disappears. It is affirmed that images with these characteristics are not included in $T$. Moreover, \#1, \#2, and \#3 are real because they can be recognized as numbers. Therefore, it can be said that these images are unsupposable-data candidates. Other images included in Fig. \ref{fig09} may also be unsupposable data candidates; however, in this study, only images \#1 to \#3 were affirmed.

\section{Evaluation and Discussion}\label{discuss}
From the results presented in Section \ref{experiment}, it was confirmed that applying the proposed UTG can generate unsupposable-data candidates. Since unsupposable-data candidates can be unsupposable data depending on the developer, it means that UTG can be useful for generating unsupposable data. By referring to the generated unsupposable data, the developer can recognize unsupposable features and create different unsupposable data having those features. Then, by testing the model by using the unsupposable data generated by UTG and the unsupposable data created by the developer, the developer can confirm the behavior of the model more strictly than hitherto possible. 

Two implementations of UTG were shown: VAE (in Section \ref{impvae}) and VQ-VAE with PixelCNN (in Section \ref{impvq}). If the target data is structured data, it is supposed that the VAE implementation is used. If the target data is image data, the image quality of the generated data is expected to be higher than that of the VAE implementation, so the VQ-VAE-with-PixelCNN implementation is preferred. In addition, VQ-VAE-2 \cite{vqvae2} has been proposed as a method of generating data with higher image quality than that possible with VQ-VAE. By implementing the proposed UTG with VQ-VAE-2, it is possible to further improve the quality of generated data.

The concept of UTG is to acquire a latent value with a low occurrence probability in prior distribution $P$ and use it to generate data. That is, the probability density function shown in Definition \ref{def01} and the algorithm shown in Algorithm \ref{al01} are examples of the method of acquiring a latent value, and it is also possible to acquire a latent value by other methods. For example, it is conceivable to modify Algorithm \ref{al01} to create an algorithm by which 
the values of $diff$ for which $d[v]$ exceeds threshold $t$ are totaled and evenly distributed only to $d[v]$ that do not exceed $t$.
Several conceivable methods were tried, and Algorithm \ref{al01} was adopted as the method for obtaining the most unsupposable data as for the MNIST dataset. Different algorithms may be suitable for other datasets.

As stated in Section \ref{intro}, data can be classified into three types: (1) supposable data, (2) unsupposable data, and (3) unreal data. 
Dataset of (1) includes existing dataset $T$ and data that can be assumed from $T$; that is, data with characteristics similar to $T$. On the contrary, dataset of (3) does not have characteristics similar to $T$. 
UTG is based on the assumption that ``In the process in which the data of (1) gradually loses the features included in $T$ and eventually changes to the data of (3), it may temporarily becomes the data of (2)." 
For example, if parameter $(\mu^{u}, \sigma^{u})$ of UTG on VAE is taken as (0, 1), the distribution given by $f$ agrees with $P$. Accordingly, most of the data included in $LU$ is supposable data. Hereafter, the higher the value of $(\mu^{u}, \sigma^{u})$, 
the lower the occurrence rate of the acquired latent value in $P$. As described in Section \ref{sec_vae}, $P$ is the distribution of (unobservable) features of the data included in $T$. Therefore, data generated from a latent value with high occurrence probability in $P$ has similar characteristics to the data included in $T$. Conversely, the lower the probability of occurrence of the latent value in $P$, the less likely it will be that the generated data will have similar characteristics to the data contained in $T$. That is, if the value of $(\mu^{u}, \sigma^{u})$ is gradually increased, the generated data changes from the data with characteristics similar to $T$ [i.e., data of (1)] to the data with no characteristics of $T$ [i.e., data of (3)]. In a similar manner, with the implementation of UTG with VQ-VAE with PixelCNN, when the value of parameter $t$ is gradually decreased from $1$, the generated data changes from (1) to (3).

\begin{figure}[htbp]
\scalebox{0.4}{\includegraphics[bb=0 0 604 220]{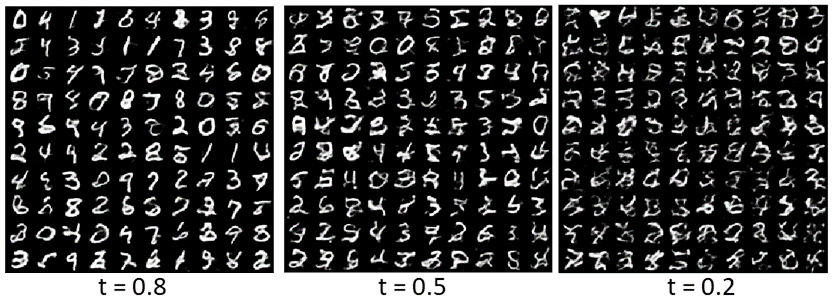}}
\caption{Change of generated data with changing $t$}
\label{fig10}
\end{figure}

UTG is based on the assumption that it is possible for data to pass through (2) in the process of changing from (1) to (3). For example, in regard to the MNIST dataset, this assumption is considered to hold. An example of the generated data when parameter $t$ is changed in the UTG implementation with VQ-VAE with PixelCNN is shown in Fig. \ref{fig10}. This figure confirms that as $t$ decreases, the data changes from (1) [supposable] to (2) [unsupposable] to (3) [unreal]. 
Whether this assumption holds or not depends on the problem domain. Data space of (1) contains data having features that make sense in the problem domain. On the contrary, data space of (3) is an area of data that does not have those features. In the process of changing from (1) to (3), meaningful features are gradually lost. UTG is effective when data with partially lost features that make sense can become unsupposable data. For example, in the case of the MNIST dataset, in the process of changing from (1) to (3), the numerical (visual) features are gradually lost, and a non-numeric image is finally generated. The image that partially lost its numerical features generated 
is not an image that does not make sense as a digit at all; it is sometimes regarded as an image of a digit that ``collapse" by handwriting. 
In this way, UTG is considered to be effective in problem domains in which if data partially loses its features, it still makes sense.

As a typical problem domain in which UTG is not very effective, the image-recognition problem of an object with a complicated occurrence can be considered. For example, living things and vehicles are composed of various combinations of characteristics, and all the characteristics are occurrence-constituent elements of the object. It is therefore highly likely that an image that has partially lost its characteristics becomes unreal data instead of unsupposable data. For example, an airplane with a missing piece of wing is no longer an airplane. From the above considerations, it is considered that the effectiveness of UTG depends on whether the meaning is completely lost when the features that constitute the meaning in the target problem domain are partially lost. Furthermore, even if the data generated by UTG is real data, whether it is unsupposable data depends on the knowledge and experience of the developer. 

\section{Conclusion}\label{conclusion}
A method called UTG---for generating unsupposable data for developers by utilizing VAE---was proposed. As for UTG, a latent value with a low occurrence probability in the prior distribution of the VAE is obtained. Unsupposable data can be generated by inputting the acquired latent value into the VAE decoder. By referring to the generated data, the developer can recognize new unsupposable features and create another unsupposable data with those features. Then, by testing the model using the unsupposable data generated by UTG and the unsupposable data created by the developer, the developer can confirm the behavior of the model more strictly than hitherto possible. 
Methods for implementing UTG on VAE and on VQ-VAE with PixelCNN were described. It was also shown that UTG can be useful for generating unsupposable data when applied to the MNIST dataset and the House Sales Price dataset. 

As for future work, UTG will be applied to other data sets to confirm its effectiveness. In particular, by applying UTG to image data such as the CIFAR-10 dataset \cite{cifar10}, it is hoped to confirm the assumption about its effectiveness discussed in Section \ref{discuss}. It will also be evaluated whether the image quality of the generated data can be further improved by implementing UTG on VQ-VAE-2.



\begin{thebibliography}{00}

\bibitem{reluplex} G. Katz, C. Barrett, D.L. Dill, K. Julian, and M.J. Kochenderfer: Reluplex: An Efficient SMT Solver for Verifying Deep Neural Networks, Computer Aided Verification 2017, pp. 97-117 (2017).

\bibitem{dlv} X. Huang, M. Kwiatkowska, S. Wang, and M. Wu: Safety Verification of Deep Neural Networks, Computer Aided Verification 2017, Lecture Notes in Computer Science, vol.10426, pp.3-29 (2017).

\bibitem{planet} R. Ehlers: Formal verification of piece-wise linear feed-forward neural networks, Automated Technology for Verification and Analysis (2017).

\bibitem{starbased} H. Tran, P. Musau, D. Manzanas Lopez, X. Yang, L. V. Nguyen, W. Xiang, and T. T. Johnson: Star-Based Reachability Analsysis for Deep Neural Networks, In 23rd International Symposisum on Formal Methods (2019).

\bibitem{xgboostveri} N. Sato, H. Kuruma, Y. Nakagawa, and H. Ogawa: Formal Verification of a Decision-tree Ensemble Model and Detection of its Violation Ranges, IEICE Transaction D, Vol.E103-D, No.02, pp.363-378 (2020).

\bibitem{deeptest} Y. Tian, K. Pei, S. Jana, and B. Ray.: DeepTest: Automated Testing of Deep-Neural-Network-driven Autonomous Cars, ICSE'2018 Technical Papers (2018).

\bibitem{deepxplore} K. Pei, Y. Cao, J. Yang, and S. Jana.: DeepXplore: Automated Whitebox Testing of Deep Learning Systems, The 26th ACM Symposium on Operating Systems Principles (2017)

\bibitem{tensorfuzz} A. Odena and I. Goodfellow: Tensorfuzz: Debugging neural networks with coverage-guided fuzzing, ICML (2019). 

\bibitem{unit} M.Y. Liu, T. Breuel, and J. Kautz: Unsupervised Imageto-image Translation Networks, In Adv. NIPS, pp.700-708 (2017).
\bibitem{cyclegan} J.-Y. Zhu, T. Park, P. Isola, A. A. Efros: Unpaired Image-to-Image Translation using Cycle-Consistent Adversarial Networks, In Proc. International Conference on Computer Vision (ICCV) (2017).

\bibitem{deeproad} M. Zhang, Y. Zhang, L. Zhang, C. Liu, and S. Khurshid: DeepRoad: GAN-Based Metamorphic Testing and Input Validation Framework for Autonomous Driving Systems, In Proc. ASE'18 (2018).

\bibitem{gan1} I. Goodfellow, J. Pouget-Abadie, M. Mirza, B. Xu, D. Warde-Farley, S. Ozair, A. Courville, and Y. Bengio:
Generative adversarial nets, Advances in Neural Information Processing Systems (2014).

\bibitem{gan2} M.-Y. Liu and O. Tuzel: Coupled generative adversarial networks, Advances in Neural Information Processing Systems (2016).

\bibitem{vae1} D. P. Kingma and M. Welling: Auto-encoding variational bayes, International Conference on Learning Representations (2014).

\bibitem{vae2} A.B.L. Larsen, S.K. Sonderby, H. Larochelle, and O. Winther: Autoencoding beyond pixels using a learned similarity metric, International Conference on Machine Learning (2016).

\bibitem{vae3} D.J. Rezende, S. Mohamed, and D. Wierstra: Stochastic backpropagation and variational inference in deep latent gaussian models, International Conference on Machine Learning (2014).

\bibitem{vqvae2} A. Razavi, A. v. d. Oord, and O. Vinyals: Generating Diverse High-Fidelity Images with VQ-VAE-2, Advances in Neural Information Processing Systems (NIPS) 32 (2019).

\bibitem{vqvae} A.v.d. Oord, O. Vinyals, and K. Kavukcuoglu: Neural discrete representation learning, CoRR, abs/1711.00937 (2017).

\bibitem{driverless} Z.Q. Zhou and L. Sun: Metamorphic Testing of Driverless Cars, Comm. ACM, vol.62, no.3, pp.61-67 (2019).

\bibitem{metamor1} T.Y. Chen, S.C. Chung, and S.M. Yiu: Metamorphic Testing - A New Approach for Generating Next Test Cases,HKUST-CS98-01, The Hong Kong University of Science and Technology (1998).

\bibitem{metamor2} T.Y. Chen, F.-C. Kuo, H. Liu, P.-L. Poon, D. Towey, Y.H. Tse, and Z.Q. Zhou: Metamorphic Testing: A Review of Challenges and Opportunities, ACM Computing Surveys, vol.51, no.1, Article No.4, pp.1-27 (2018).

\bibitem{adversarial} I.J. Goodfellow, J. Shelens, and C. Szegedy: Explaining and Harnessing Adversarial Examples, ICRL2015, arXive:1412.6572 (2014).

\bibitem{metropolis} I. Beichl and F. Sullivan: The Metropolis Algorithm, in Computing in Science \& Engineering, vol.2, no.1, pp.65-69 (2000).

\bibitem{mnist} Y. LeCun, C. Cortes, and C. J. C. Burges. The MNIST database of handwritten digits. 
http://yann.lecun.com/exdb/mnist/. (Accessed on April 28th, 2020)

\bibitem{cifar10} A. Krizhevsky: The CIFAR-10 dataset, CIFAR-10 and CIFAR-100 datasets. 
https://www.cs.toronto.edu/~kriz/cifar.html (Accessed on April 28th, 2020) 






\end{thebibliography}
\end{document}